%% file: main.tex
\documentclass[sigconf,9pt]{acmart}
\usepackage[nolist,nohyperlinks]{acronym} %approved
\usepackage{graphicx} %approved
\AtBeginDocument{%
  \providecommand\BibTeX{{%
    \normalfont B\kern-0.5em{\scshape i\kern-0.25em b}\kern-0.8em\TeX}}}

\acmYear{2022}\copyrightyear{2022}
\setcopyright{rightsretained}
\acmConference[ACM MobiCom '22]{The 28th Annual International Conference On Mobile Computing And Networking}{October 17--21, 2022}{Sydney, NSW, Australia}
\acmBooktitle{The 28th Annual International Conference On Mobile Computing And Networking (ACM MobiCom '22), October 17--21, 2022, Sydney, NSW, Australia}
\acmPrice{}
\acmDOI{10.1145/3495243.3560525}
\acmISBN{978-1-4503-9181-8/22/10}
\graphicspath{{fig/}}

\begin{document}

\input{acronyms.tex}
% https://capitalizemytitle.com/
\title{AdaptOver: Adaptive Overshadowing Attacks in Cellular Networks}

\author{Simon Erni}
\email{simon.erni@inf.ethz.ch}
\orcid{0000-0002-0740-701X}
\affiliation{%
  \institution{ETH Zurich}
  \country{Switzerland}
}

\author{Martin Kotuliak}
\email{martin.kotuliak@inf.ethz.ch}
\orcid{0000-0002-5833-4447}
\affiliation{%
  \institution{ETH Zurich}
  \country{Switzerland}
}

\author{Patrick Leu}
\email{patrick.leu@inf.ethz.ch}
\orcid{0000-0003-0069-3778}
\affiliation{%
  \institution{ETH Zurich}
  \country{Switzerland}
}

\author{Marc Roeschlin}
\email{marc.roeschlin@inf.ethz.ch}
\orcid{0000-0003-0519-9066}
\affiliation{%
  \institution{ETH Zurich}
  \country{Switzerland}
}

\author{Srdjan Capkun}
\email{srdjan.capkun@inf.ethz.ch}
\orcid{0000-0001-5727-3673}
\affiliation{%
  \institution{ETH Zurich}
  \country{Switzerland}
}

% Short List:
\renewcommand{\shortauthors}{Erni et al.}

\begin{abstract}
\input{sections/00_abstract}
\end{abstract}

%! suppress = EscapeUnderscore


\maketitle

\input{sections/01_introduction}

\input{sections/02_background}
\input{sections/03_adaptover}
\input{sections/04_implementation.tex}
\input{sections/05_evaluation}
\input{sections/06_discussion}

\input{sections/07_related_work}
\input{sections/08_conclusion}

\begin{acks}
\input{sections/91_acks}
\end{acks}

\bibliographystyle{ACM-Reference-Format}
%\input{output.bbl}
\input{refs.bbl}

\appendix

\end{document}

%% file: acronyms.tex
% !TeX root = main.tex

\acrodef{mac}[MAC]{Medium Access Control}
\acrodef{TDMA}[TDMA]{Time Division Multiple Access}
\acrodef{WPAN}[WPAN]{Wireless Personal Area Network}
\acrodef{GSM}[GSM]{Global System for Mobile communication}
\acrodef{IEEE}[IEEE]{Institute of Electrical and Electronics Engineers}
\acrodef{tls}[TLS]{Transport Layer Security}
\acrodef{mnc}[MNC]{Mobile Network Code}
\acrodef{fbs}[FBS]{Fake Base Station}
\acrodef{dci}[DCI]{Downlink Control Information}
\acrodef{rf}[RF]{Radio Frequency}
\acrodef{crs}[CRS]{Cell-Specific Reference Signals}
\acrodef{lte}[LTE]{Long-Term Evolution}
\acrodef{pcfich}[PCFICH]{Physical Control Format Indicator Channel}
\acrodef{evm}[EVM]{Error Vector Magnitude}
\acrodef{ue}[UE]{User Equipment}
\acrodef{enb}[eNodeB]{Evolved Node B}
\acrodef{mib}[MIB]{Master Information Block}
\acrodef{sib}[SIB]{System Information Block}
\acrodef{eps}[EPS]{Evolved Packet System}
\acrodef{emm}[EMM]{EPS Mobility Management}
\acrodef{mme}[MME]{Mobility Management Entity}
\acrodef{imsi}[IMSI]{International Mobile Subscriber Identity}
\acrodef{tmsi}[TMSI]{Temporary Mobile Subscriber Identity}
\acrodef{cs}[CS]{Circuit Switched}
\acrodef{ps}[PS]{Packet Switched}
\acrodef{sdr}[SDR]{Software Defined Radio}
\acrodef{rrc}[RRC]{Radio Resource Control}
\acrodef{imei}[IMEI]{International Mobile Equipment Identity}
\acrodef{stmsi}[S-TMSI]{Serving Temporary Mobile Subscriber Identity}
\acrodef{nas}[NAS]{Non-Access Stratum}
\acrodef{epc}[EPC]{Evolved Packet Core}
\acrodef{pdcch}[PDCCH]{Physical Downlink Control Channel}
\acrodef{pdsch}[PDSCH]{Physical Downlink Shared Channel}
\acrodef{usrp}[USRP]{Universal Software Radio Peripheral}
\acrodef{gpsdo}[GPSDO]{GPS Disciplined Oscillator}
\acrodef{gps}[GPS]{Global Positioning System}
\acrodef{PSS}[PSS]{Primary Synchronization Signal}
\acrodef{SSS}[SSS]{Secondary Synchronization Signal}
\acrodef{rnti}[RNTI]{Radio Network Temporary Identifier}
\acrodef{sirnti}[SI-RNTI]{System Information RNTI}
\acrodef{drx}[DRX]{Discontinuous Reception}
\acrodef{dos}[DoS]{Denial of Service}
\acrodef{plmn}[PLMN]{Public Land Mobile Network}
\acrodef{asn1}[ASN.1]{Abstract Syntax Notation One}
\acrodef{prach}[PRACH]{Physical Random Access Channel}
\acrodef{guti}[GUTI]{Globally Unique Temporary Identifier}
\acrodef{usim}[USIM]{Universal Subscriber Identity Module}
\acrodef{ofdm}[OFDM]{Orthogonal Frequency Division Multiplexing}
\acrodef{ofdma}[OFDMA]{Orthogonal Frequency Division Multiple Access}
\acrodef{rms}[RMS]{Root Mean Square}
\acrodef{qam}[QAM]{Quadrature amplitude modulation}
\acrodef{fdd}[FDD]{Frequency Division Duplex}
\acrodef{tdd}[TDD]{Time Division Duplexing}
\acrodef{scfdma}[SC-FDMA]{Single-carrier Frequency Division Multiple Access}
\acrodef{papr}[PAPR]{Peak to Average Power Ratio}
\acrodef{ran}[RAN]{Radio Access Network}
\acrodef{mno}[MNO]{Mobile Network Operator}
\acrodef{harq}[HARQ]{Hybrid Automatic Repeat Request}
\acrodef{lcid}[LCID]{Logical Channel ID}
\acrodef{ccch}[CCCH]{Common Control Channel}
\acrodef{dcch}[DCCH]{Dedicated Control Channel}
\acrodef{dtch}[DTCH]{Dedicated Traffic Channel}
\acrodef{rlc}[RLC]{Radio Link Control}
\acrodef{pdcp}[PDCP]{Packet Data Convergence Protocol}
\acrodef{phich}[PHICH]{Physical HARQ Indicator Channel}
\acrodef{sic}[SIC]{Successive Interference Cancellation}
\acrodef{tpc}[TPC]{Transmit Power Control}
\acrodef{ta}[TA]{Timing Advance}
\acrodef{rar}[RAR]{Random Access Response}
\acrodef{suci}[SUCI]{Subscription Concealed Identifier}

%% file: sections/00_abstract.tex
% !TeX spellcheck = en_US

In cellular networks, attacks on the communication link between a mobile device and the core network significantly impact privacy and availability. Up until now, fake base stations have been required to execute such attacks. Since they require a continuously high output power to attract victims, they are limited in range and can be easily detected both by operators and dedicated apps on users' smartphones.

This paper introduces AdaptOver---a MITM attack system designed for cellular networks, specifically for LTE and 5G-NSA. AdaptOver allows an adversary to decode, overshadow (replace) and inject arbitrary messages over the air in either direction between the network and the mobile device. Using overshadowing, AdaptOver can cause a persistent ($\geq$ 12h) DoS or a privacy leak by triggering a UE to transmit its persistent identifier (IMSI) in plain text. These attacks can be launched against all users within a cell or specifically target a victim based on its phone number.

We implement AdaptOver using a software-defined radio and a low-cost amplification setup. We demonstrate the effects and practicality of the attacks on a live operational LTE and 5G-NSA network with a wide range of smartphones. Our experiments show that AdaptOver can launch an attack on a victim more than 3.8km away from the attacker. Given its practicability and efficiency, AdaptOver shows that existing countermeasures that are focused on fake base stations are no longer sufficient, marking a paradigm shift for designing security mechanisms in cellular networks.

%% file: sections/01_introduction.tex
% !TeX spellcheck = en_US
\section{Introduction}
\label{sec:introduction}

Cellular networks, such as LTE and 5G, provide wide-area connectivity even in challenging environments. They are designed to cope with interference but are not designed to be jamming-resilient. Therefore, it is not surprising that the radio link between the base station and the \acf{ue} is vulnerable to jamming attacks~\cite{lichtman_ltelte-jamming_2016}.
In addition to jamming attacks, cellular networks are vulnerable to a range of attacks by fake base stations. In those attacks, the UEs attempt to attach to a fake base station, and as a result, an attacker, playing a Man-in-the-Middle (MITM) role, is able to spoof all messages that are not integrity-protected. Such manipulations result in \ac{dos} attacks~\cite{jover_lte_2016, shaik_practical_2016, hussain_lteinspector_2018} or cause privacy issues by leaking the permanent identity of the UE (\ac{imsi})~\cite{jover_lte_2016, basin_formal_2018}. Although both wireless jamming and fake base station attacks can have a high impact, they require the attacker to remain active for long periods of time and to use high output power, risking detection by operators \cite{nakarmi_detecting_2018}, law enforcement agencies \cite{esd_esd_2022, rohdeschwarz_gmbh__co_kg_rsnestor_2022}, and even individuals \cite{cellularprivacy_cellularprivacyandroid-imsi-catcher-detector_2020, srlabs_snoopsnitch_2019, echeverria_phoenix_2021, li_fbs-radar_2017, ney_seaglass_2017, quintin_detecting_2020}; there is a number of commercial tools available to detect such attacks.

In contrast to jamming and fake base station attacks, signal overshadowing attacks aim to disrupt cellular networks by replacing legitimate wireless signals over the air. 

Overshadowing attacks require a precisely time- and frequency- synchronized transmission with signal strengths slightly stronger (+3dB, \cite{yang_hiding_2019}) or even weaker (-3.4dB,  \cite{ludant_sigunder_2021}) than the legitimate signal. Existing  \ac{dos} overshadowing attacks \cite{yang_hiding_2019, ludant_sigunder_2021} modify the downlink broadcast messages of a cell and can mark it as unavailable. However, in many environments, the UE can choose from many nearby cells and will immediately hop to the next available one. To effectively cause a DoS, an attacker must therefore overshadow all available cells simultaneously. Furthermore, base stations are typically mounted in elevated places and output a high power signal on many bands simultaneously, which in practice requires the attacker to transmit with significant power to achieve a large attack range. Finally, continuous high-power attacks increase the likelihood of the attacks being detected.

In this work, we introduce AdaptOver, a MITM system based on \emph{adaptive overshadowing}. AdaptOver introduces the modification of higher layer protocols and is the first framework that can overshadow both downlink and uplink. Using uplink overshadowing, AdaptOver can modify messages from UEs anywhere within a cell, regardless of the position of the UE or the attacker.

Since AdaptOver executes higher layer (\ac{nas}) protocol attacks, it can modify ongoing procedures between the UE and the core network in real-time. This marks a paradigm shift in the attacker model of cellular networks since such attacks were believed to require a fake base station. Our results show that by overshadowing a small set of \ac{nas} messages on either the downlink or uplink, AdaptOver is able to introduce stealthy and persistent ($\geq$12 hours long) DoS, or privacy leakage by making the \ac{ue} broadcast its \ac{imsi} in plain-text. Since the UE does not re-try on a different cell immediately, AdaptOver can launch both an effective and persistent DoS attack even when operating on a single cell.

While AdaptOver can attack all UEs in a cell in parallel, it can also attack UEs selectively based on their \ac{tmsi}. The \ac{tmsi} can be obtained automatically from a phone number with an intersection attack, requiring only a device capable of sending \textit{silent SMS}.

We evaluated AdaptOver on 20 different smartphone models from 7 different vendors containing 17 different basebands. We evaluated our uplink overshadowing attacks in urban and rural environments and achieved an attack range of more than 3.8km in a rural and 398m in an urban environment. In comparison, downlink overshadowing attacks are less efficient; we found that downlink overshadowing achieves a range of 50m in ideal LOS conditions with the same attacker setup.

In summary, we make the following contributions:

\begin{itemize} 

\item We develop a reactive and protocol-aware MITM attack system, enabling the injection of uplink and downlink messages on the radio link at any point in time and on any communication layer of LTE.
\item We demonstrate the impact of AdaptOver by implementing DoS attacks that are stealthier than prior attacks (which rely on fake base stations), are more persistent (12h), more practical (impact all base stations), and have significantly more range (the whole cell coverage area), than existing overshadowing attacks.
\item We do an extensive evaluation of both uplink and downlink overshadowing attacks in a real-world deployment and show that uplink and downlink attacks on cellular networks are practical. 
\item Finally, we discuss countermeasures that baseband manufacturers and operators could implement to mitigate the proposed DoS attacks.

\end{itemize}

%% file: sections/02_background.tex
% !TeX spellcheck = en_US
\section{Background}
\label{sec:background}

This section presents the background knowledge required to understand the design of the AdaptOver attack.

\subsection{Components of 4G \& 5G-NSA}
An LTE network consists of two main components, the core network, called Evolved Packet Core (EPC), and the Radio Access Network (RAN), which includes a set of base stations called \ac{enb} and user equipment (UE) devices, such as smartphones, routers, or IoT devices.

In 5G-NSA (non-stand-alone), the components are the same as in LTE with the addition of 5G base stations (gNodeB) operating the NR wireless protocol. 5G-NSA is a preliminary step to a full 5G network, where the control-plane messages between UEs and network core are still sent exclusively through LTE base stations. Using 5G-NSA, a UE maintains parallel connections to multiple cells and radio technologies. 5G base stations serve as secondary cells, transporting data-plane messages only. In 5G-SA (stand-alone) networks, control plane messages will be also be routed through gNodeBs. In this paper, we mainly look into LTE and 5G-NSA protocol-level attacks; specifically, we attack UEs using the control-plane messages sent over LTE. Therefore, the rest of the section discusses procedures and identifiers in LTE.

On the physical layer (PHY), the wireless signal of LTE consists of subframes of 1ms duration. Each subframe in LTE has the same basic structure and is independent from other subframes.

\subsection{Identifiers}
Every UE is identified by a unique persistent identifier, called \ac{imsi}. This identifier is sent in clear text the very first time a UE connects to the network. Otherwise, to hide the users' identity, a temporary identifier \ac{tmsi} is used. The network can update the TMSI at any time. However, not all operators update the TMSI frequently~\cite{hussain_privacy_2019}.

Each message sent on the downlink or uplink of a particular base station is identified with a short-term identifier RNTI. An RNTI can address a particular UE directly or is reserved for a specific purpose (e.g., broadcast system configuration, paging, or random access).

\subsection{Procedures}
\begin{figure}
	\centering
	\includegraphics[width=0.9\linewidth,page=1]{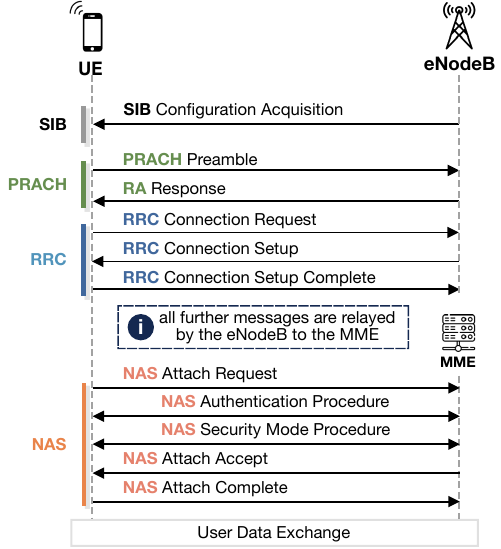}
	\caption{Full attachment procedure of a UE to a network}
	\label{fig:attachment_procedure}
\end{figure}
When a UE attaches to a network, it executes the procedure as shown in Figure~\ref{fig:attachment_procedure}. Regardless if the UE has been merely idle or if it performs a new connection establishment, it first attaches to the base station. After this, the UE continues to start the core network attachment procedure, which will ultimately grant the UE network and cellular services.
\subsubsection{Base Station Attachment} After acquiring the cell configuration by decoding the \ac{sib}, the UE requests an uplink allocation by transmitting a PRACH Preamble. Next, the eNodeB allocates an \ac{rnti} and the required uplink resources and signals this to the UE with the \ac{rar}. The UE then initiates an RRC Connection Request containing the establishment cause and the identifier. The identifier can be either a TMSI (if previously allocated to the UE) or a random number. The eNodeB then reflects the entire request in the subsequent downlink message (RRC Connection Setup) acting as a contention resolution, i.e., preventing two UEs using the same \ac{rnti}. Moreover, the RRC Connection Setup message contains dedicated configuration for the UE, such as signaling power levels and RLC channel configuration. Finally, if the contention resolution ID matches the RRC Connection Request, the UE confirms the connection using the RRC Connection Setup Complete message.

\subsubsection{Core Network Attachment} After having established a connection with a base station using the RRC procedure described above, the UE starts the \ac{nas} attachment procedure. First, the UE sends its identifier (TMSI if available, else IMSI) with the NAS Attach Request to the MME. The MME and the UE will then perform the authentication and key exchange (AKA) procedure, followed by activating encryption and integrity protection of the communication in the security mode procedure. Finally, both parties (UE and MME) conclude the attachment with a NAS Attach Accept and Complete message. Whenever the UE and the network do not exchange data for some time, the UE enters idle mode. After some time, it may resume the connection by repeating the base station attachment procedure, followed by a NAS Service Request, which is integrity protected using the previous session key. If it can be verified correctly, the Security Mode Procedure is done, after which user data exchange can resume.

All of the steps executed during the core network attachment can result in a rejection by the network core. Causes for a reject differ, and the behavior of the UE upon receipt of a reject message also varies, as specified in 3GPP TS 24.301 \cite{3gpp_3gpp_2020}. With some, the UE re-attaches right away (e.g., Service Reject \#9 for failed integrity check of the Service Request), whereas with others, it disconnects and will not try to attach for more than 12 hours (e.g., Attach Reject \#8, all LTE \& non-LTE services are forbidden)). The user can make the phone re-try earlier by toggling flight mode, re-inserting the SIM card, or restarting the phone. Attacks using such messages originating from a fake base station were publicized by Jover in \cite{jover_lte_2016}. 3GPP updated the specification 24.301 in version V13.5.0, specifying that when the UE receives a non-integrity protected reject message, it should re-try to attach within 30-60mins (T3247) up to a certain limit. In our tests, only two phones implemented this.

%% file: sections/03_adaptover.tex
% !TeX spellcheck = en_US
\section{AdaptOver}
\label{sec:adaptover}

AdaptOver is a MITM attack built for 4G/5G-NSA cellular networks. It implements adaptive downlink sniffing and overshadowing on both downlink and uplink, allowing an attacker to inject messages on the \ac{nas} layer. In this section, we describe the high-level operation of AdaptOver by showing examples of attacks using AdaptOver that cause DoS and privacy leakage on cellular networks. Implementation details of AdaptOver are given in Section~\ref{sec:implementation}.

Our attacks target the UE attach and service resumption procedures, which allow the injection of unauthenticated messages. We show how surgical manipulations of these procedures lead to UE IMSI disclosure and persistent DoS on UEs. To achieve this, AdaptOver first synchronizes with the message exchange between UE and eNodeB, then decodes downlink messages, and finally modifies (overshadows) either downlink or uplink messages, depending on the specific attack.

Downlink overshadowing directly changes the messages that the UE receives. Uplink overshadowing, on the other hand, changes the message(s) from the UE to the eNodeB, requiring only a little more power than a regular UE (i.e., >23dBm). The modified message then triggers a response from the eNodeB / MME that aligns with the attacker's goal. This response is then transmitted using the legitimate eNodeB itself, impacting the whole cell coverage area at once. Consequently, AdaptOver also evades all fake base station detection mechanisms, which were previously used to detect such protocol attacks. Finally, since both downlink and uplink attacks are based on the 3GPP standard, the attacks are portable and executable on any operator.

An attacker can use AdaptOver to attack all connections in a cell. Alternatively, e.g., using only a phone number as identification, AdaptOver can fully automatically launch a targeted attack against a victim. The attacks shown in our work are not meant to be an exhaustive list of what AdaptOver is capable of. For example, we implemented variants of the DoS attacks on the down- and uplink that influence the authentication procedure to result in an authentication reject. Since the outcome is the same ($\geq$ 12h DoS), we did not include an analysis of all such (sub-)variants in this paper.

\subsection{Attack Assumptions}
In general, we assume that the attacker has no access to any cryptographic material and is not able to compromise any part of the cellular network infrastructure or user equipment. The attacker only operates on the wireless channel.

We also assume that the attacker is located within or in close proximity of the (targeted) cell, such that decoding the downlink is possible.
Similarly, the attacker can receive and decode the messages transmitted by the user equipment of the victim.
Furthermore, in order to overshadow a transmission, the attacker needs to be able to transmit with enough power such that the attack signal is 3dB stronger than the legitimate signal at the victim's receiver (or at the base station in case of uplink overshadowing). 
Both receiving and transmitting with high enough power can be achieved by placing the attacker's devices in proximity of the victim/base station, leveraging high-gain (directional) antennas, or using amplification.
Since the transmit power of a UE is regulated in the LTE standard and depends on the distance to the base station, an attacker will find it easier to overshadow messages on the uplink for a specific UE than the downlink. In addition, base stations often use power output levels on the order of 50 watts or higher.

The attacker deploys commercial off-the-shelf (COTS) hardware to decode communication and inject messages. We do not assume any specialized hardware or low latency processing that goes beyond the available COTS hardware, such as commercially available software-defined radios.

We further consider a stronger variant of the attacker described above, which has all the listed capabilities and, on top of that, managed to acquire the victim's phone number, e.g., through a social engineering attack.

\subsection{Connection Information Gathering}\label{sec:prepare}
\begin{figure}
	\centering
	\includegraphics[width=\linewidth,page=2]{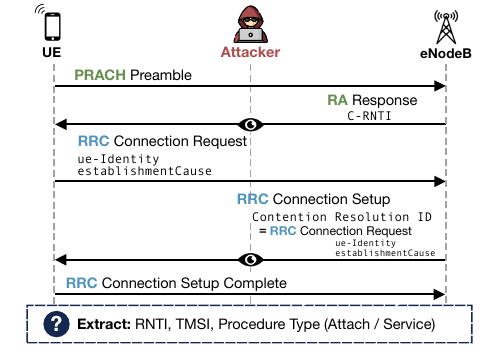}
	\caption{Initial connection procedure of a UE, annotated with which information elements AdaptOver can extract from it}
	\label{fig:attack_adaption}
\end{figure}

A distinctive feature of AdaptOver is its adaption to each UE connection. AdaptOver achieves this by listening on the downlink. Figure~\ref{fig:attack_adaption} highlights the most important individual messages exchanged between UE and \ac{enb}. AdaptOver first listens to \texttt{RA Responses} and thus knows the C-RNTI of all current connections. Next, the UE sends an RRC Connection Request. Depending on the current procedure, the \texttt{ue-Identity} field contains a random number (Attach Request), or the \ac{tmsi} (Service Request). The \texttt{establishment Cause} field further indicates the type of connection procedure that will follow. The eNodeB accepts this request with an \texttt{RRC Connection Setup} response. In the MAC layer, a \texttt{Contention Resolution ID} is transported along with the response. This ID is a 1:1 encoding of the RRC Connection Request, which means that after AdaptOver receives it, it learns the \ac{tmsi} and procedure type of the connection and can decide whether and how to launch the attack. Finally, the \texttt{RRC Connection Setup} message contains information on the physical-layer encoding of further messages, which AdaptOver needs to use to generate the overshadowing signal correctly.

\subsection{AdaptOver: Targeted Attacks }\label{ref:tmsi_intersection}
With AdaptOver, an attacker can launch large-scale attacks by targeting a large number of UE connections or can specifically target UEs, identified by their phone number or \ac{tmsi}. Since the \ac{tmsi} is a temporary identifier, the challenge for the attacker is to obtain the \ac{tmsi}. To this end, AdaptOver leverages ideas from~\cite{hussain_privacy_2019}. Since we assume that the attacker knows the victim's phone number, AdaptOver can send a series of silent SMS (Short Message Type 0) to the UE. Silent SMS are completely stealthy, do not trigger any notification at the UE of the victim, and can be sent automatically, e.g., with the Huawei E3372 LTE USB Stick, exposing a AT Command interface. The SMS are sent with a random delay between 20s and 30s, such that each SMS generates a paging message. AdaptOver receives all paging messages on the downlink and checks which \ac{tmsi}s were in paging messages received shortly after it sent the SMS. By checking which \ac{tmsi} was received most often with the least variance of the paging delay, AdaptOver automatically determines which \ac{tmsi} belongs to the victim. Even in a busy tracking area, with hundreds of paging messages per second (we measured > 350 per second in an urban setting), AdaptOver automatically determines the \ac{tmsi} with <10 SMS messages.

\subsection{AdaptOver: Denial of Service (DoS)}

\subsubsection{AdaptOver: Downlink DoS} 
\begin{figure}
	\centering
	\includegraphics[width=\linewidth,page=3]{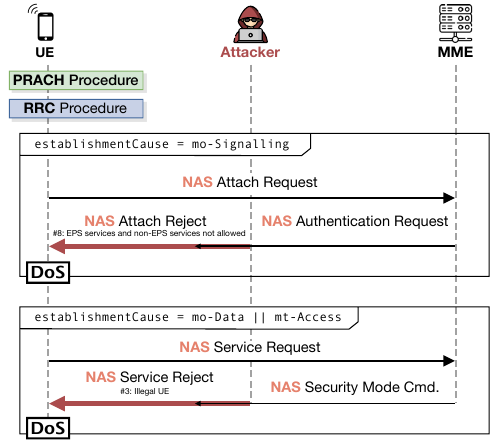}
	\caption{AdaptOver Downlink DoS Attack resulting in a Service or Attach Reject}
	\label{fig:dos_dl}
\end{figure}

The attacker first records the downlink during connection establishment between the UE and the eNodeB as described in Section \ref{sec:prepare} and learns the type of connection procedure that will follow. If it is an attach procedure, the attacker will overshadow the response of the MME to an attach request (e.g., an \texttt{Authentication Request}) with an \texttt{Attach Reject}. AdaptOver overshadows the downlink with this message continuously for 250ms, in order to transmit all necessary acknowledgments (see Section~\ref{sec:reliable_transport_modification} for details) and to make sure that the legitimate response is not received, as it is impossible to predict when exactly the MME / eNodeB sends it exactly. After the attack, the UE enters a state where it will not attempt to re-connect to any other cell of the operator in the same tracking area for either 30-60mins or more than 12 hours, depending on the model of the phone. If it is a service request procedure (i.e., a connection re-establishment), the attacker will react with a \texttt{Service Reject}, overshadowing the \texttt{Security Command} that the MME usually sends. In this case, AdaptOver only needs to overshadow for 50ms (see Section~\ref{sec:reliable_transport_modification} for details), since only one acknowledgment needs to be transmitted. Figure \ref{fig:dos_dl} summarizes the attack. Note that the eNodeB is omitted from the figure.

\subsubsection{AdaptOver: Uplink DoS}
AdaptOver first learns from the adaptive phase which connection procedure the UE will use. In most cases, this is going to be a service request procedure, in which case the \texttt{Service Request} sent by the UE will be overshadowed with one containing an invalid MAC in the short message authentication code field of the \ac{nas} header. According to TS24.301, this will prompt the MME to reply with a \texttt{Service Reject} \#9, prompting the UE to start an attach procedure to immediately re-connect. Note that in this re-attachment, the UE will use a random connection identifier; thus, AdaptOver is not able to link it to the previously used TMSI with 100\% confidence.

\begin{figure}
	\centering
	\includegraphics[width=\linewidth,page=4]{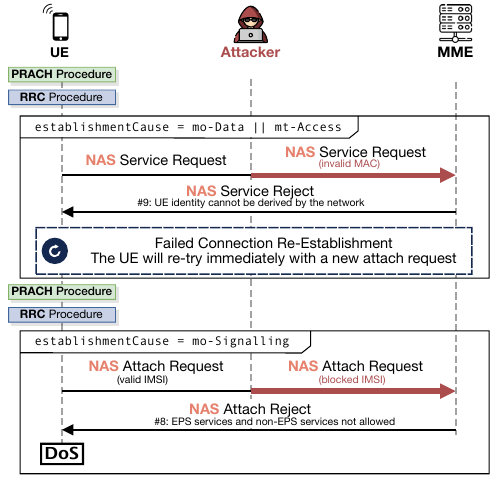}
	\caption{AdaptOver Uplink DoS Attack resulting in an Attach Reject}
	\label{fig:dos_ul}
\end{figure}

If AdaptOver recognizes the connection to be an attachment procedure, it will overshadow the \texttt{Attach Request} sent by the UE with one containing an IMSI that is blocked in the network, resulting in an \texttt{Attach Reject} with cause value \# 8. Upon receiving the \texttt{Attach Reject} \#8, the UE will enter a DoS state and not try to re-connect for $\geq$ 12 hours. Figure \ref{fig:dos_ul} shows the attack procedure. 

For this attack to work, the IMSI sent by the attacker in the \texttt{Attach Request} must trigger the \texttt{Attach Reject} with a cause value corresponding to a persistent DoS (e.g., \#8, but \#15 also works well). This IMSI may belong to a blocked or invalid SIM and can be found by the attacker brute-forcing the IMSI space or reading out the IMSI of an expired old SIM. Finally, it should be noted that any IMSI of any network worldwide can be tried; thus, the search space is enormous. In our experiments, we found such a set of blocked IMSIs for reject codes \#8 and \#9 with less than 100 trials.

Note that while the DoS impact of the response messages (i.e., \texttt{Attach / Service Reject}) are well-studied, they previously required the use of a fake base station to actually inject the message. In our case, AdaptOver turns the legitimate base station into an attacker's asset, leveraging it's power amplification, transmission range, and reliable transport mechanisms to be able to attack all connecting UEs with minimal effort.

\subsection{AdaptOver: Uplink IMSI Extractor}\label{sec:privacy}

In LTE networks, IMSI catchers are a dire problem. However, until AdaptOver was introduced, IMSI catchers required the attacker to operate a fake base station, luring the victim devices to connect to it by transmitting with high power on high-priority frequency bands. This means that countermeasures against IMSI catchers could focus on the presence of fake base stations and their identifiable characteristics (e.g., signal strength, unusual parameters).

With AdaptOver and its principle of surgical message overshadowing, this is no longer sufficient. LTrack~\cite{kotuliak_ltrack_2021} introduces \textit{IMSI Extractor}, a downlink IMSI catcher based on a previous version of AdaptOver. In that work, the authors leverage AdaptOver to inject an \texttt{Identity Request} via downlink overshadowing, prompting the UE to reply on the uplink with its \ac{imsi} embedded within the \texttt{Identity Response}. AdaptOver has the distinct advantage that it is able to inject messages into an ongoing connection, removing many assumptions that fake base station detectors are based on.

In this paper, we introduce a further improved IMSI catcher attack leveraging uplink overshadowing, which we name Uplink IMSI Extractor. Since uplink overshadowing works from any point within a cell with only little power, AdaptOver is able to cover the whole cell with this attack. This works reliably even in dense urban environments.

\begin{figure}
	\centering
	\includegraphics[width=\linewidth,page=5]{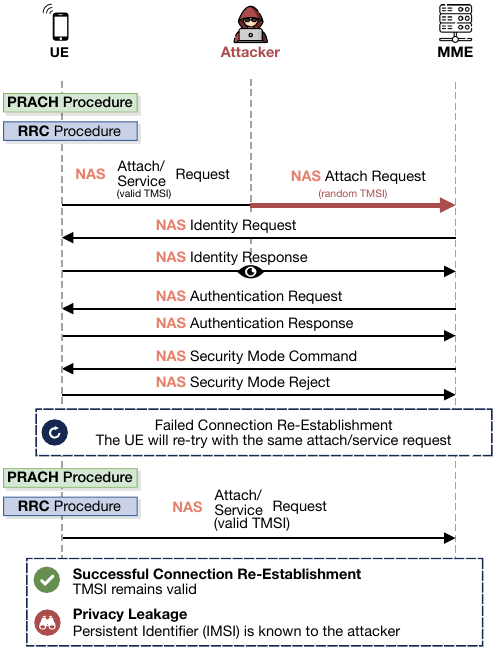}
	\caption{AdaptOver Uplink IMSI Extractor Attack}
	\label{fig:id_ul}
\end{figure}

Our attack is shown in Figure~\ref{fig:id_ul}. AdaptOver overshadows the \texttt{Attach / Service Request} with an \texttt{Attach Request} containing a random \ac{tmsi} that is unknown to the MME. As the MME is unable to link the connection attempt to a previous UE context, it will initiate an identification procedure, starting with an \texttt{Identity Request}. The UE will respond to that with its IMSI in plaintext, embedded in the \texttt{Identity Response}, ready for the attacker to capture.

Given the nature of this attack, we still need to assume that the attacker is able to receive the IMSI sent by the UE in return. For that, the attacker has an uplink receiver, preferably in proximity to the (static) base station. The decoding of the uplink can then happen later with arbitrary latency. This means that an attacker can also just record and store IQ samples of the uplink band with a small and cheap setup and analyze it offline to decode the IMSI.

After the attack, the \ac{tmsi} used by the UE remains the same. As shown in Figure \ref{fig:id_ul}, the procedure will continue with the authentication procedure but will fail during the security mode procedure. This is because, in the \texttt{Security Mode Command}, the network replays the received UE capabilities received in the \texttt{Attach Request}. Since AdaptOver overshadowed that \texttt{Attach Request} with capabilities not matching the one of the UE (i.e., all set to false), the UE will react with \texttt{Security Mode Reject} and will abort the entire connection attempt. However, the UE will immediately re-try the original (re-)attachment procedure, leaving the whole context, including the \ac{tmsi}, intact. AdaptOver will then no longer attack the connection attempts of the attacked \ac{tmsi}(s). This way, AdaptOver allows the attacker to link both previous and future passive connection observations containing only the \ac{tmsi} together with the persistent identifier \ac{imsi}.

%% file: sections/04_implementation.tex
% !TeX spellcheck = en_US
\section{Implementation}

This section presents the implementation of uplink and downlink AdaptOver. Our implementation is based on the open-source library srsRAN~\cite{puschmann_srsltesrslte_2020}. The implementations for the two overshadowing directions differ significantly. Therefore, we describe them independently.

\label{sec:implementation}
\subsection{Downlink Overshadowing}
AdaptOver's downlink implementation consists of the following components:
\begin{enumerate}
    \item The real-time downlink decoder continuously decodes the physical and higher-layer messages sent by the base station. It is used to establish the precise time and frequency offset necessary for overshadowing, to configure the encoding and parameters of the messages, and to trigger the start of the attack. Its implementation is similar to a UE's downlink decoder.
    \item After the attack is triggered, the Message Encoding component encodes and packages the attack messages.
    \item To allow the UE to send a request during overshadowing, fake uplink allocations are injected.
    \item Control messages are injected continuously that modify the reliable transport mechanisms. This is done so that the victim UE successfully transmits its request. Otherwise, it will not accept the attacker's response.
    \item Messages are then transmitted with an \ac{sdr}, overshadowing the downlink by precisely aligning the time and frequency of the output signal of the attacker.
\end{enumerate}

\subsubsection{Real-Time Downlink Decoder}\label{sec:adaptover_dl_decoder}
AdaptOver's real-time downlink decoder synchronizes to the base station continuously and decodes the connection establishment procedure for each connecting \ac{ue}. AdaptOver first listens for \ac{rar} messages from the base station. These messages allocate the \ac{rnti} to the UE. For each \ac{rnti} allocation, each subsequent subframe is tested if it contains data destined to the RNTI by blind-decoding the \ac{pdcch} for \ac{dci} messages. After the downlink decoder receives the \texttt{RRC Connection Setup} message, the attack is triggered, and the \ac{rnti} is ignored again, such as to free decoding resources. The decoding task itself is distributed across a pool of threads and is optimized to minimize copying operations. This design enables the decoding of many concurrent connection establishment procedures. We tested this with a Intel(R) Xeon(R) Gold 6242 CPU, which can decode up to 50 RNTIs in parallel, which is sufficient even for busy urban cells. In our experiments with busy real cells, we found that several performance optimizations were necessary to help the system operate reliably. One of the most helpful optimizations is to isolate (\texttt{isolcpus}) and pin one CPU core to only receive samples from the SDR, as this is the most latency-sensitive task. This optimization prevents buffer overflows, where the radio is unable to transport enough samples in time to the CPU.

\subsubsection{Message Encoding}\label{sec:message_encoding} After the attack is triggered, AdaptOver encodes the desired attack messages from the NAS procedure level down to the physical layer. Using the obtained configuration messages, AdaptOver configures each channel and layer the message passes through because the configuration can differ for every UE connection. First, the NAS message is generated (i.e., \texttt{Attach / Service Reject}) and packed as a payload into the \texttt{dedicatedInfoNAS} field of the \texttt{RRC Connection Setup Complete} message. This RRC message is then passed to the PDCP layer, where the sequence number (starting with 0 for the first message) and the message authentication tag are appended to the message. Since no keys are activated during the attack, this field contains only zeroes. Next, the message is passed to the RLC layer, appending another sequence number starting at 0. The MAC layer multiplexes the attack message together with the RLC acknowledgement (see Section~\ref{sec:reliable_transport_modification}). Finally, the message is passed to the transmission queue, where it is ready to be combined with attack messages destined for other victims in parallel. The encoding down to IQ samples to be transmitted via the SDR is done just-in-time with a buffer of 2 subframes to reduce the reaction time of the attacker.

\subsubsection{Fake Uplink Allocations}\label{sec:fake_uplink_allocatins} In LTE, the base station allocates resource slots to the UEs, indicating when and how they may send their data. Without such uplink allocations, no data can be sent by the UE. For example, the UE must be able to send the Service Request message before it can accept the Service Reject response. But as all original allocations from the eNodeB are overshadowed by the attacker, the attacker needs to send its own fake uplink allocations. AdaptOver sends uplink allocations at the first subframe of every frame. It uses an internal buffer to store the sent allocations, as the \ac{harq} acknowledgments that follow depend on it. 

\subsubsection{Reliable Transport Modification}\label{sec:reliable_transport_modification} In addition to being able to send data with uplink allocations, the UE needs to have its request fully transmitted and acknowledged before it accepts a response to it. Because the attacker also overshadows acknowledgments for the uplink data, the attacker needs to include acknowledgments in the overshadowing. In LTE, acknowledgments are carried in 2 distinct layers; \ac{rlc} and \ac{mac}. 

The \ac{rlc} layer uses acknowledgments with sequence numbers in both up and downlink directions. The first segment of every connection has the sequence number 0. A segment with sequence number $x$ is acknowledged with a RLC acknowledgment with sequence number $x+1$. Suppose the current link quality does not allow encoding a message in a single uplink subframe. In that case, the \ac{rlc} layer may split the message into smaller segments, which are then sent over multiple subframes and reassembled at the receiver. Therefore, an \ac{rlc} acknowledgment must be sent to the UE for all segments of the message.
During the attack, AdaptOver sends acknowledgments for all sequence numbers in increasing order from the interval $[1, \Delta]$, where $\Delta$ is the maximum amount of segments the message might be split up in. In our experiments, we discovered that for the \texttt{Attach Reject} attack, $\Delta \leq 4$ suffices for all tested smartphone models and signaling conditions. For the \texttt{Service Reject} attack, $\Delta = 1$ is sufficient since the \texttt{Service Request} itself is very small. During the attack, the acknowledgment sequence number is incremented by one every 50ms and sent at every subframe together with the attack message combined in the \ac{mac} layer.

In the \ac{mac} layer, acknowledgments for the previous transport block sent by the UE are sent exactly 8 subframes after the uplink allocation. As explained in Section~\ref{sec:fake_uplink_allocatins}, the attacker sends the acknowledgments at every frame on subframe 8, after having sent an uplink allocation in subframe 0. \ac{harq} acknowledgments depend on the location of the corresponding uplink acknowledgment in the LTE resource grid, which is looked up in the buffer created in Section~\ref{sec:fake_uplink_allocatins}.

\subsubsection{Downlink Transmission} Overshadowing requires precise frequency and time synchronization. Using the realtime downlink decoder, AdaptOver can continuously measure and adjust its time and frequency offset to the real base station before every transmission burst. In addition, AdaptOver includes reference signals on every subframe, allowing the channel estimation at the receiver to tune reliably to the attacker's channel. Second, each \ac{pdsch} transmission to the UE is modulated adaptively according to the configuration messages decoded from the downlink just before starting the attack. Third, multiple messages may be transmitted simultaneously on the \ac{pdsch} channel, enabling parallel attacks on multiple victim UEs. Finally, high-level messages are encoded less than 2ms before transmitting them to the radio device, which enables a highly reactive system as the downlink decoder and message encoding component can place messages in the send buffer until 2ms before the intended transmission time. 

\subsubsection{Power Management}
For downlink overshadowing to work, the transmission must be received with higher power at the UE. Therefore, the attacker must use amplification and high gain antennas to achieve a high range. Furthermore, the location of the transmitter must be chosen adequately since the legitimate base stations are placed in locations with good RF propagation properties as well. We used the same amplification setup for both downlink and uplink overshadowing and compared their range in Section \ref{sec:evaluation}.

\subsection{Uplink Overshadowing}
The uplink overshadowing implementation is closely based on srsUE by srsRAN~\cite{puschmann_srsltesrslte_2020}. Remarkably, for a proof-of-concept implementation of uplink overshadowing working in a lab environment, we only needed changes in 14 lines of code in srsUE, showcasing the technical feasibility of the attack. The main idea behind the implementation of uplink overshadowing in AdaptOver is for it to act as a UE with a few modifications:
\begin{enumerate}
    \item A real-time downlink decoder constantly monitors the downlink for new UE connections and decodes the physical layer for the UEs under attack. It uses the same approach as described in Section~\ref{sec:adaptover_dl_decoder}.
    \item For each new UE connection, a separate attack-UE implementation is spawned, implementing the attack-specific behavior. The higher layer processing is done with the same approach as outlined in Section~\ref{sec:message_encoding}.
    \item The transmissions of all attack-UEs are combined and sent aligned in time and frequency with the legitimate UE transmission with as little additional transmit power as possible.
\end{enumerate}

\subsubsection{Downlink Decoding}
The base station schedules the transmissions of all UEs on the uplink. It sends time and frequency resource allocations to the UEs, enabling multiple UEs to transmit data simultaneously without interfering. The base station then decodes the resources it previously allocated to the UE. Consequently, AdaptOver must decode and buffer all uplink resource allocations destined for the victim. Moreover, to allow selective attacks on UEs based on their TMSI, AdaptOver tracks and decodes the UE connection until it receives a \texttt{RRC Connection Setup} accompanied by a \texttt{Contention Resolution} in the MAC layer. Using the TMSI and the cause of the connection inside the \texttt{RRC Connection Setup}, AdaptOver can then decide whether to attack the UE or not.
\subsubsection{Reliable and Large-Scale Overshadowing}

One of the main challenges for uplink overshadowing is supporting a real cell with many connections in parallel. For this scenario, having a modified version of a UE is not enough. Therefore, we modified the architecture of srsUE to run multiple higher-layer UE stacks on top of a single instance of the physical layer implementation. The physical layer camps on one cell and listens to new connections announced via \ac{rar} messages. Once received, it assigns the new connection to one higher-layer stack, which changes its state to RRC Connected. The stack then prepares the uplink payload and sends it to the physical layer, which transmits it in an allocation previously observed on the downlink. After the attack is completed, we reset the stack by changing its state to RRC Idle and do not engage it until a new \ac{rar} message is available. During a running attack, other received \ac{rar} messages and UE connections can still be handled by other stacks, creating a reliable and large-scale attack mechanism. Using this architecture, we can simulate multiple AdaptOver or even regular UE instances running in parallel using just one SDR.

During our testing, AdaptOver was able to cover all the UEs in a base station in a high-density area in the middle of the city with just four higher-layer stacks on a computer with an Intel i9-10900KF processor. If needed, it can support at least ten stacks in parallel.

\subsubsection{Power Management}

As explained in the previous subsection, the attacker and the legitimate UE are transmitting simultaneously on the same frequency. Because of the capture effect, only the one received with the higher power is decoded by the base station. The transmit power of the legitimate UE in the \ac{prach} procedure is controlled in an open-loop by indicating the target receive power in \ac{sib} messages. After this, the base station actively controls the transmit power of the UE by sending \ac{tpc} commands in the \ac{mac} layer. UEs closer to the base station send with low power to conserve power and lower emissions. UEs far away will send with higher power, up to a maximum defined by 3GPP and local regulations, set in \ac{sib} messages. Because of this power control loop, AdaptOver is able to overshadow a legitimate UE anywhere in a cell, even from a remote location, since the required power to overshadow at the base station stays largely constant. Finally, since the power output of an SDR is limited (e.g., a B210 has <10dBm), amplification is necessary to overshadow the UE if the attacker is not very close to the base station. In our experiments, we used a cheap commercially available amplification and filtering setup, costing around 200 USD in total (without the SDR). To determine the required transmission power for a given attack location, the attacker can either measure the received power of the base station and add the power margin corresponding to the estimated path loss or can alternatively increase the transmit gain until the attack works reliably.

\subsubsection{Precise Synchronization}

To align the reception times of multiple UE transmissions, the base station instructs the UE to send its data earlier or later using \ac{ta} commands. The resulting \ac{ta} value is a function of the distance between the regular UE and the base station. AdaptOver is designed to ignore all \ac{ta} commands and applies a static timing advancement established in advance of the attack. If the location of the base station is known, the attacker can determine its position, distance and thus the resulting timing advancement. If it is not known, the attacker can connect to the base station once with srsUE and read the \ac{ta} value sent in the \ac{rar} and apply this value to all its future transmissions.

\subsubsection{Reliable Transport Discussion}
In contrast to downlink overshadowing, no active reliable transport manipulation needs to be done. Uplink allocations and uplink acknowledgments are both carried on the downlink and left intact, so there is no need to modify them. For uplink allocations, it could be that after AdaptOver is done sending, the UE still has not managed to transmit its request in full. In this case, the UE will ask for more uplink allocations and transmit the rest of the data, getting each subframe acknowledged by the base station. This data will be discarded by the base station since a complete message with the same RLC message sequence number has already been received from AdaptOver. If AdaptOver sends more RLC segments than the regular UE, the base station will send more RLC acknowledgments to the UE, which will discard them, similarly to downlink overshadowing.

%% file: sections/05_evaluation.tex
% !TeX spellcheck = en_US

\section{Experimental Evaluation}
\label{sec:evaluation}

We structure the evaluation of AdaptOver in multiple parts. First, we evaluate the reaction of the UEs and the operators to the intended attack messages to verify that they indeed cause the intended \ac{dos} or privacy leakage. We then evaluate the attacks on a real operator network in various channel conditions. Finally, we look into publicly available attack detection apps and test if they detect AdaptOver.

\subsection{Attacker Setup}
\subsubsection{Lab Experiments}
To verify the UE behavior, we set up a private LTE and 5G-NSA network using Amarisoft Callbox \cite{amarisoft_amari_2022}, a shielded box \cite{mos_equipment_mission_2022} containing the UEs under test, and a USRP B210 as the attacker. This setup allows us to closely monitor and record the connection logs during our attacks, as well as debug and develop the attacks before trying them on the live network. We also had access to the Faraday cage of the operator to evaluate our attacks. There, real base stations connected to the production core network of the operator were available to us. We did not need to use any amplification for the attacks to work in these environments.

\subsubsection{Real World Experiments}\label{sec:setup_real_world}
\begin{figure}
	\centering
	\includegraphics[width=\linewidth,page=6]{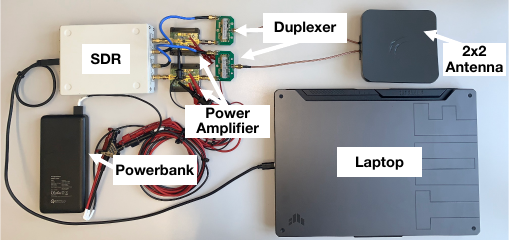}
	\caption{AdaptOver Attacker Setup}
	\label{fig:attack_setup}
\end{figure}

As the attacker, we always used a USRP B210 software-defined radio connected over USB 3.0 to a laptop with an Intel Core i7-11800H CPU. To amplify the signals, we used two Qorvo amplifier evaluation boards (TQP9111-PCB2600), connected to a 5V power supply. Filtering was done using a dielectric duplexer \cite{laurent_dielectric_2022} for the respective band under attack. Finally, we used the Mikrotik mAnt LTE 5o 5dBi antenna \cite{mikrotik_mikrotik_2022}. Figure~\ref{fig:attack_setup} shows the setup for the uplink attacker. For the downlink attacker, the same setup was used, but without duplexers since the transmit and receive frequencies are the same for a downlink attacker. In this case, a second 2x2 antenna (same model) was connected to the receiver port of the SDR.

\subsubsection{Limiting Impact to other customers}\label{sec:limiting_customerimpact}
In real-world experiments, real customers must not be impacted. In order to guarantee this, we limit the attacks carried out in the real world to only target UEs that are under our control. AdaptOver determines the TMSI of our UEs with the intersection attack before running the targeted overshadowing attacks. Thus, we can be sure that if we have our current TMSI, no other customer will be actively attacked. To verify the correctness of the intersection attack, we extracted the session key between the SIM and the UE using SimTrace2 \cite{welte_simtrace_2021} and verified the PDCP MAC of the downlink messages in the connection identified by the TMSI. Since there are no TMSIs in the RRC Connection during an Attach Request, we executed attacks only while the phone is in the \texttt{Service Request} procedure to eliminate the possibility of attacking other users.

Still, in the case of downlink overshadowing, there is collateral damage in the form of service degradation to other customers during the attack since all reference signals of the downlink are overshadowed. This is why we tested downlink overshadowing attacks only in a shielded environment or on an deserted open field.

\subsection{DoS Behavior of UEs \& Operators}
In Table~\ref{table:attack_results}, we summarized all 20 smartphone models tested and their reaction to the attacks. The behavior of the UEs in reaction to DoS attacks is specified in 3GPP TS 24.301 \cite{3gpp_3gpp_2020}, and we verified it using downlink overshadowing with real base stations in the operators' Faraday cage. In our shielded lab environment, we measured the time during which the phone did not try to re-connect to the network. Note that only the Attach Reject attack is presented since the Service Reject attack had a matching impact. Some of the newer UEs we tested implemented T3247, as specified in Section 5.3.7b of TS 24.301 \cite{3gpp_3gpp_2020}. This timer indicates to the UE to re-try within 30-60 minutes until a UE-specific maximum number of failed attach attempts is reached. When this number of attempts is reached, all smartphones tested exhibited a $\geq$ 12h DoS when subjected to an \texttt{Attach Reject \#8}. Note that a user interacting with the device has the possibility to restore connection earlier, e.g., via rebooting the phone or re-inserting the SIM card. Without such actions, after the DoS attack, the phone did not try to use any other cell of the same operator. By attacking only one cell at a time, AdaptOver can cause an effective and persistent DoS.

After having verified the behavior of each UE, we verified that the operator in our country sends the desired attack messages when prompted. For this, we connected with a slightly modified version of srsUE. We verified that when the network receives a \texttt{Service Request} without authentication, it reacts with a \texttt{Service Reject \#9}, as is expected by the TS24.301 standard. Furthermore, using random IMSIs, we found a set of IMSIs that were blocked in the real network and yielded an \texttt{Attach Reject \#8} when used in an \texttt{Attach Request}. We also found that all other invalid IMSIs trigger the \texttt{Attach Reject \#15}, which has a similiar persistent DoS effect. Finally, we verified that upon receiving an \texttt{Attach Request} with an invalid TMSI, the network reacts with an \texttt{Identity Request}, which is used in the uplink IMSI extraction attack.

\subsection{Downlink Overshadowing Range}
\begin{figure}
	\centering
	\includegraphics[width=\linewidth,page=7]{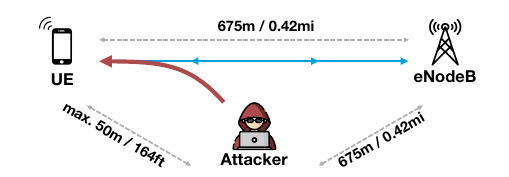}
	\caption{AdaptOver Downlink Range}
	\label{fig:attack_range_dl}
\end{figure}

We evaluated the range of AdaptOver downlink overshadowing using the amplification setup as shown in Section~\ref{sec:setup_real_world}. In order to limit disturbance of the real network during our attack as a side effect of downlink overshadowing, we went to an urban environment on an open field, with LOS conditions between UE, eNodeB, and attacker. We started the downlink DoS overshadowing attack and repeatedly connected an LG Nexus 5X to the commercial network. During that, we gradually increased the distance between the attacker and the UE while keeping the distance to the base station always the same. After a distance of 50m / 164ft, the attack did no longer work. The output power (before the antenna) for this setup was 29dBm at each port.

\subsection{Uplink Overshadowing on Real Phones}
We connected each phone to the commercial network and sent a series of silent SMS to the UE to extract its TMSI. Table~\ref{table:attack_results} shows that none of the UEs displayed any notification after having received these SMS. Next, we launched the IMSI Extractor attack and verified that we received an Identity Request on the downlink. Finally, we launched the uplink DoS attack, triggering an Attach Reject on the downlink. This worked for all phones in Table~\ref{table:attack_results}.

\subsection{Uplink Overshadowing Range}
\begin{figure}
	\centering
	\includegraphics[width=\linewidth,page=8]{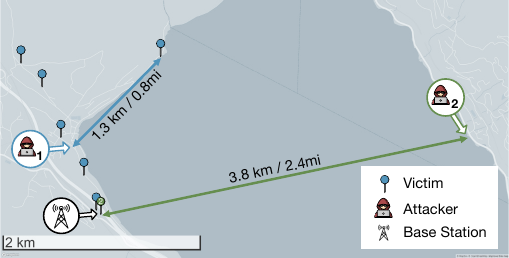}
	\caption{AdaptOver Real World Evaluation in a rural environment}
	\label{fig:attack_rural}
\end{figure}

After having verified that we can overshadow every UE on the uplink on the real network, we moved on to establish a practical range of the attack. For this, we first went to a rural and urban environment. In the rural environment, we set up the attacker at a static location (1) with LOS to the eNodeB, and moved around with the test UE (LG Nexus 5X), marking on Figure~\ref{fig:attack_rural} with (1) where the attack worked. We achieved a maximum distance of 1.3km between the attacker and the UE, and the attack still worked even if the UE was right next to the base station. Finally, we did a long-distance experiment (marked with 2), where we moved the attacker across the lake, leaving the test UE right next to the base station. Even in this challenging setup, the attack worked reliably. The output power (before the antenna) for this setup was 14.2dBm at each port, with a timing advance of 26.04$\mu s$.

\begin{figure}
	\centering
	\includegraphics[width=\linewidth,page=9]{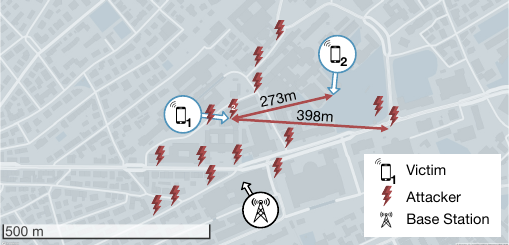}
	\caption{AdaptOver Real World Evaluation in an urban environment. Lightning bolts denote attacker locations for two different experiments/victims (1 and 2).}
	\label{fig:attack_urban}
\end{figure}

In the rural environment, most of the time both the UE and the attacker were in LOS of the eNodeB. For urban environments, this does not hold necessarily, as the attacker might be unable to establish LOS to the desired cell. We connected a Samsung Galaxy S21 as 5G-NSA capable UE at a static location indoors (1), with the connection indicator showing 5G. Shown in Figure~\ref{fig:attack_urban} with (1) is where uplink overshadowing using AdaptOver worked. We were able to move around in the whole coverage area of the cell, and the attack worked even at the very edge. As long as we were able to receive and decode the downlink of the cell, the attack then always worked. Most of the attack locations were in NLOS conditions. The maximum distance we achieved between the victim and the UE was 398m / 0.2mi. We also moved the UE to a location (2), 273m away from the attacker (2), and the attack worked in this setup as well. The transmit power in this environment was at most 28.5dBm at each port with a timing advance of 5.73$\mu s$.

\begin{table}
	\centering
	\caption{Summary of AdaptOver Attacks for 20 phone models from 7 vendors with 17 distinct basebands}
		    	\label{table:attack_results}
		\begin{tabular}{@{}lccc}
		    \toprule
		                     &  \textbf{DoS Time}   & \textbf{Silent} &   \textbf{Uplink}    \\
		                     &   \textbf{Attach}    &  \textbf{SMS}   & \textbf{DoS \& IMSI} \\
		    \textbf{Phone}   & \textbf{ Reject \#8} & \textbf{hidden} &  \textbf{Extractor}  \\ \midrule
		    Huawei P20 Pro   &        > 12h         &      \checkmark       &         \checkmark         \\
		    Huawei P30       &        > 12h         &      \checkmark       &         \checkmark         \\
		    Huawei P30 Lite  &        > 12h         &      \checkmark       &         \checkmark         \\
		    Huawei P40 5G    &        > 12h         &      \checkmark       &         \checkmark         \\
		    Samsung A8       &        > 12h         &      \checkmark       &         \checkmark         \\
		    Samsung S10      &        > 12h         &      \checkmark       &         \checkmark         \\
		    Samsung S21 5G   &        > 12h         &      \checkmark       &         \checkmark         \\
		    LG Nexus 5X      &        > 12h         &      \checkmark       &         \checkmark         \\
		    iPhone 6S        &        > 12h         &      \checkmark       &         \checkmark         \\
		    iPhone 7         &        > 12h         &      \checkmark       &         \checkmark         \\
		    iPhone 8         &        > 12h         &      \checkmark       &         \checkmark         \\
		    iPhone 11        &        > 12h         &      \checkmark       &         \checkmark         \\
		    iPhone X         &        > 12h         &      \checkmark       &         \checkmark         \\
		    Xiaomi Mi 9      &        > 12h         &      \checkmark       &         \checkmark         \\
		    Xiaomi Mix 3 5G  &        > 12h         &      \checkmark       &         \checkmark         \\
		    Pixel 2          &        > 12h         &      \checkmark       &         \checkmark         \\
		    Pixel 3a         &        > 12h         &      \checkmark       &         \checkmark         \\
		    Pixel 4          &        > 12h         &      \checkmark       &         \checkmark         \\
		    Pixel 5 5G       &     2x 30s, >12h     &      \checkmark       &         \checkmark         \\
		    OnePlus 9 Pro 5G &     5x 30-60min,     &      \checkmark       &         \checkmark         \\
		                     &     10x 10s,>12h     &                 &                      \\ \bottomrule
		\end{tabular}

\end{table}

\subsection{Impact on other phones}
AdaptOver can execute a targeted attack against a specific UE based on its TMSI present in the \texttt{RRC Connection Request}. However, a downlink overshadowing attack overshadows every subframe continuously for 50ms during a \texttt{Service Reject} attack, and for 250ms during a \texttt{Attach Reject} attack. This is why other users that are not attacked are still affected. On the other hand, for an uplink overshadowing attack, we expect no interference with other users since AdaptOver only transmits on resource blocks that were already assigned to the victim UE by the base station.

\begin{figure}
	\centering
	\includegraphics[width=\linewidth]{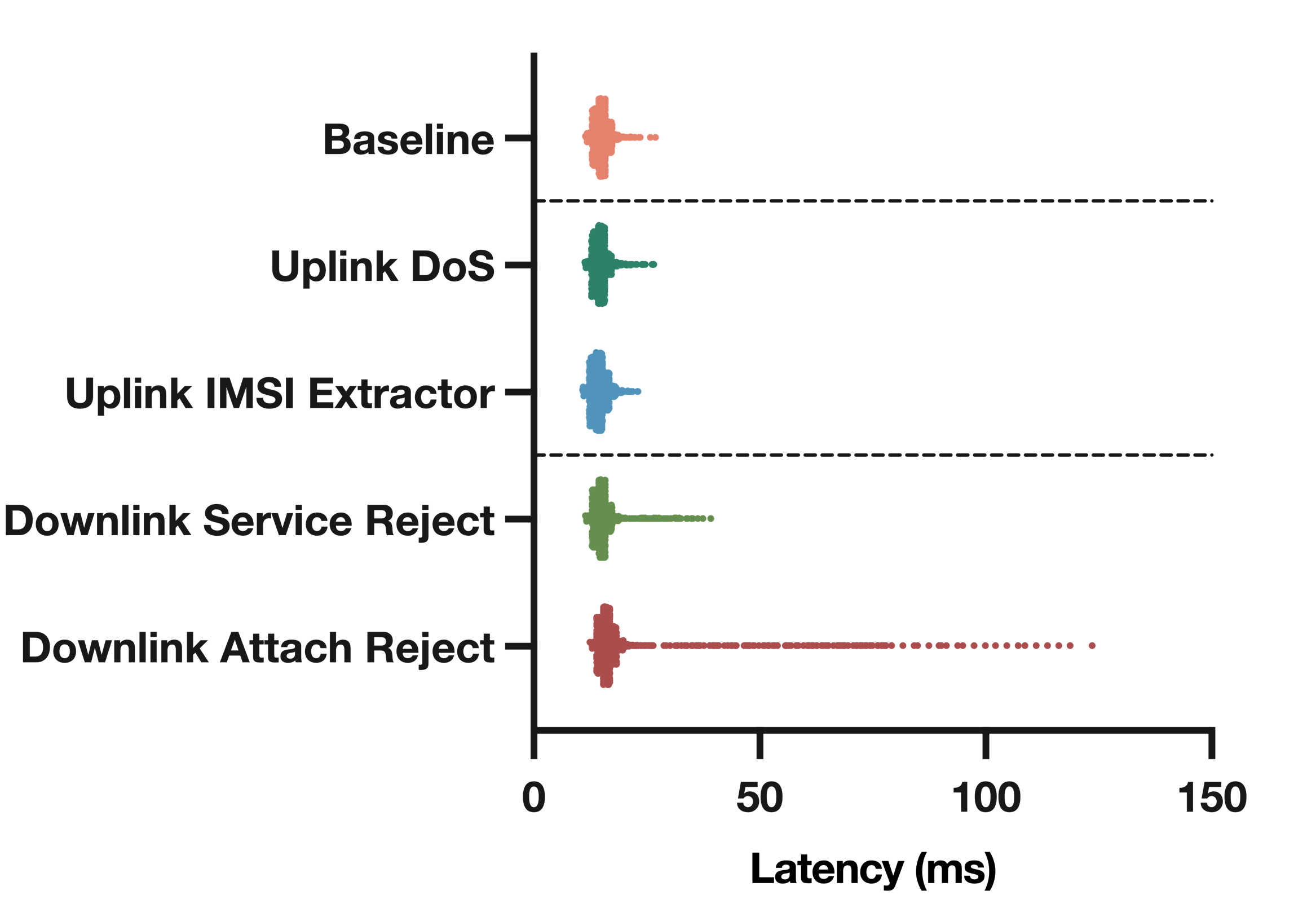}
	\caption{Latency as measured from a non-attacked UE}
	\label{fig:impact_other_phones}
\end{figure}

We evaluated this aspect as follows. We had two computers running a continuous sockperf \cite{ivanov_mellanoxsockperf_2022} latency measurement via our private Amarisoft LTE network over an LTE router (Mikrotik SXT R). We measured the latency between the sockperf client and server for 10 seconds in various conditions. For the baseline scenario, we had another UE resume service during this 10-second window. For the attack scenarios, we executed the attack once during the 10-second window on another UE. Sockperf was run in the \textit{under-load} mode with 200 packets and replies per second. We repeated each scenario three times and, for each scenario, recorded the latency of every packet, which we present in the scatter plot in Figure~\ref{fig:impact_other_phones}.

The baseline and the uplink attacks have a very similiar latency distribution, therefore an uplink overshadowing attack on one UE does not impact other UEs. However, during the downlink attacks, an increased latency during a few points is visible, especially for the longer-lasting attach reject attack.

\subsection{IMSI Catcher Detection Apps}
We installed CellularPrivacy~\cite{cellularprivacy_cellularprivacyandroid-imsi-catcher-detector_2020} on a Huawei P40 and ran the Uplink IMSI Extractor attack, which resulted in no detection. We then tried SnoopSnitch~\cite{srlabs_snoopsnitch_2019} on a rooted LG Nexus 5X, which also did not detect the attack. To exclude experimental errors, we also examined their source code and did find that their detection algorithms are not suited to detect AdaptOver attacks.

%% file: sections/06_discussion.tex
% !TeX spellcheck = en_US

\section{Discussion}
\label{sec:discussion}

\subsection{Disclosure and Ethical Concerns}
In the development and evaluation of AdaptOver, we were in close collaboration with an operator and their equipment provider, shared our attack procedures and results, and acquired approval to run the experiments on the real network. As shown in Section~\ref{sec:limiting_customerimpact}, we can exclude any customer impact due to our experiments.

\subsection{LTE Attack Methods in Comparison}\label{ref:compare_lte_attacks}

We compare three existing attack methods on cellular networks against AdaptOver; jamming, fake base stations, and existing overshadowing attacks. Jamming cellular signals requires the continuous presence of the attacker and needs large amounts of energy on all of the bands where the operator transmits. Fake Base Stations are also able to attack higher layers in the communication but also require constant high energy transmission and are detectable with numerous methods. Downlink overshadowing attacks as introduced by SigOver~\cite{yang_hiding_2019} are more efficient and stealthier than fake base station attacks. The authors implemented a DoS attack using IMSI paging and SIB overshadowing. SigOver is considered an eye-opener since it clearly documents the low requirements in terms of adversarial power. Downlink overshadowing on the physical layer can be improved further if the precise resource allocation and content of the signal is known a priori, as demonstrated by SigUnder~\cite{ludant_sigunder_2021}. However, the SigUnder approach does not directly transfer to the overshadowing of dedicated downlink channels as required by AdaptOver, since those channels are inherently unpredictable in time, resource grid location, and content.

Compared with pure downlink overshadowing attacks like SigOver and LTrack, AdaptOver can launch attacks on the uplink. With the same power amplification, uplink attacks have the highest range, the lowest transmission time, and the lowest impact on other non-targeted UEs. We summarize the attacks' differences in Table \ref{table:attack_comparison}.

\begin{table}
	\centering
	\caption{Comparison of Overshadowing Attacks using the same attacker setup}
		    	\label{table:attack_comparison}
		\begin{tabular}{@{}lccc}
		    \toprule
		                                       & \textbf{Attack} & \textbf{Transmission} & \textbf{Impact on} \\
		    \textbf{Attack}                    & \textbf{Range}  &     \textbf{Time}     & \textbf{other UEs} \\ \midrule
		    SigOver (barred SIB)  &       50m       &      continuous       &    Complete DoS    \\
		    LTrack &       50m       &         250ms         &    Incr. latency    \\
		    AdaptOver Downlink                 &       50m       &       50-250ms        &    Incr. latency    \\
		    AdaptOver Uplink                   &      3.8km      &        11x 1ms        &        None
		\end{tabular}

\end{table}

\subsection{Potential Countermeasures}
The presented attacks integrate the adversarial transmission into the legitimate time-frequency grid and overshadow only relevant parts. Approaches for detecting fake base stations based on broadcast signals, e.g., based on synchronization signal strength or configuration information, are therefore not directly applicable since those are not affected. Similarly, signing broadcast messages as proposed in \cite{hussain_insecure_2019, singla_look_2021,3gpp_3gpp_2022} will not prevent an AdaptOver attack.

The impact of the DoS attacks could be reduced by UEs re-attaching more frequently. This does not prevent a repeated attack but would increase the load on the attacker. Indeed, in case a UE receives a service or attach reject message \emph{without} integrity protection, 3GPP specifies to wait 30-60min (T3247) before a reconnection attempt~\cite{3gpp_3gpp_2020}. Most of the tested phones do not implement this timer.

In 5G, the permanent identifier is encrypted before transmission on the uplink, which prevents the IMSI extractor attack. However, two registration requests can be linked together, as implemented in~\cite{chlosta_5g_2021}. The attack, which until now uses a fake base station, can in the future also be implemented using AdaptOver. This means that fixes to the AKA protocol, such as \cite{wang_privacy-preserving_2021}, should be implemented.

Since the attacker transmits and is not assumed to be capable of annihilating signals, some physical-layer trace remains. In general, the attacker does not have the same channel as the victim and overshadows a benign transmission. These properties could be used by physical-layer detection techniques. Such an approach might measure the temporal consistency of channel state information obtained from reference signals and detect increases in overall signal power or look into the behavior of the transport layer protocol to spot obvious modifications. \ac{sic} has been proposed as a countermeasure against signal overshadowing~\cite{ludant_sigunder_2021,cover_broadcast_1972}. In order to detect and separate overshadowed signals, the receiver first decodes the message, resamples it, and subtracts it from the (buffered) incoming signal. The receiver then proceeds to decode the resulting signal again. If another message is detected, it concludes that it is subject to an overshadowing attack.
\ac{sic} might prove effective if the adversarial signal has amplitude and frequency information different enough from the overshadowed signal.

The work in~\cite{echeverria_phoenix_2021} does not propose the use of lower-layer indicators to detect the presence of a fake base station but rather examines the protocol trace of the interaction between the user equipment and (fake) base station. The proposed rules, however, would flag any rejection or identification procedure by the network as malicious, even if it might be legitimately necessary to do either one of them. Instead, we propose to harden the network core. If the network would never send rejection messages with persistent cause values in the first place, uplink overshadowing would lose its advantage. Since such countermeasures can be deployed by the operators themselves, they protect all of their customers without any action on their part. For protecting against the IMSI extractor attack, the operator could verify that the incoming RRC connection and the subsequent NAS request message have matching establishment causes since AdaptOver injects an \texttt{Attach Request} where a \texttt{Service Request} would be. Finally, \texttt{Service Request} messages with invalid signature should be ignored instead of treated with a \texttt{Service Reject}, as this presents another DoS avenue.

Finally, a base station could trick the attacker into launching the attack without a legitimate UE present, giving away the presence of the attacker. It could do so by transmitting a fake UE resource allocation and RCC connection setup message. The uplink attacker would then immediately react to this, announcing its presence.% to the base station.

\subsection{Implications \& Future Work}

Overshadowing attacks are inherent to almost all wireless communication systems. In a sense, these attacks are similar to wired network attacks where we assume the Dolev-Yao attacker model. Our work showed that similar assumptions should be taken into account when designing future cellular protocols. 
In 5G-SA systems, AdaptOver will still be a problem since overshadowing is feasible~\cite{ludant_sigunder_2021} and numerous attacks (DoS, tracking) on the NAS layer remain \cite{hu_systematic_2019}. While in 5G-SA the \ac{imsi} is encrypted and only transmitted as \ac{suci}, an attacker can still link 2 sessions by replacing the \ac{suci} of a current registration with a previously observed one \cite{basin_formal_2018, chlosta_5g_2021}. Since srsRAN, the basis of AdaptOver, is 5G-SA capable, we expect that AdaptOver can be ported with minimal adjustments to 5G-SA and that DoS and tracking attacks can be launched. In our lab 5G-SA network, we tested the Huawei P40 and the OnePlus 9 Pro, and they exhibited a >12h DoS when they received a \texttt{Service Reject \#3}.
In the meantime, many IoT devices rely on the use of cellular networks. After being subjected to an AdaptOver DoS attack applications with remote sensors could require manual intervention to regain connection, and remotely controlled distributed energy grids (i.e., solar or wind farms) can lose their controllability. In general, overshadowing can also be applied to other IoT protocols. In \cite{tan_breaking_2022}, the authors showed that NB-IoT and Cat-M protocols are susceptible to uplink and downlink overshadowing and executed attacks on the MAC, RLC, and RRC layer. This opens an opportunity to investigate whether AdaptOver's \ac{nas}-layer attacks can be applied to those protocols.
Finally, simultaneous overshadowing and reception would enable porting more fake base station attacks to AdaptOver, but it requires the attacker to be located close to the base station and good isolation between the sending and receiving part, e.g., via a circulator.

%% file: sections/07_related_work.tex
% !TeX spellcheck = en_US

\section{Related Work}
\label{sec:related_work}

A selection of protocol flaws of LTE and 5G can be found in~\cite{rupprecht_putting_2016,kim_touching_2019,basin_formal_2018} with a comprehensive overview of LTE security in~\cite{forsberg_lte_2012}. In our work, we focus on attacks that can be executed on the air interface between the base station and the UE. 

A fake base station acting as a MITM can almost arbitrarily modify user traffic on the fly, as shown in~\cite{rupprecht_imp4gt_2020, rupprecht_breaking_2019}. They have been successfully combined with protocol attacks on the NAS layer to cause DoS, as discovered and reported in~\cite{shaik_practical_2016, jover_lte_2016, hussain_lteinspector_2018}. In addition, many techniques to track users have been proposed, with the majority of attacks based on the linkability of identifiers (IMSI, TMSI, SUCI). Some of the first reported issues with IMSI paging and linkability can be found in~\cite{arapinis_new_2012}, and more recently in ~\cite{fei_lte_2019,palama_imsi_2021}, including a SUCI catcher for 5G networks~\cite{chlosta_5g_2021}. An extensive survey of commercial IMSI catchers can be found in~\cite{park_anatomy_2019}.
Even though fake base station attacks are potent, they leave significant footprint (>30dB) in the spectrum. Measures to detect them have been evaluated in numerous studies, such as~\cite{cellularprivacy_cellularprivacyandroid-imsi-catcher-detector_2020, quintin_detecting_2020, dabrowski_imsi-catch_2014, ney_seaglass_2017, dabrowski_messenger_2016, li_fbs-radar_2017, berry_detecting_2015, nakarmi_detecting_2018}. Most of them work by detecting an unusual broadcast configuration, location and  physical-layer indicators (i.e., signal strength).

Downlink overshadowing for cellular networks has first been studied in SigOver \cite{yang_hiding_2019} and SigUnder \cite{ludant_sigunder_2021}. Both of these works statically pre-generate samples and repeatedly inject them afterward. In the case of SigUnder, the attacker needs perfect knowledge of the existing resource blocks that are overshadowed. These issues make the attacks infeasible for higher layer adaptive attacks and limit it to overshadowing only the broadcast channels.

LTrack~\cite{kotuliak_ltrack_2021} presents and evaluates a covert IMSI extractor, enabling persistent UE tracking. The authors leverage a previous version of AdaptOver and combine downlink overshadowing with a downlink and uplink sniffer in order to extract the IMSI. After having acquired the IMSI, the UE can be localized and tracked using a set of passive sniffers, which record the TDOA of the uplink.

In contrast to \cite{yang_hiding_2019, ludant_sigunder_2021, kotuliak_ltrack_2021}, AdaptOver implements higher layer downlink and uplink overshadowing and, to the best of our knowledge, is the first work to demonstrate that overshadowing leads to significantly more persistent and efficient DoS and IMSI extraction attacks, especially when done on the uplink.

%% file: sections/08_conclusion.tex
% !TeX spellcheck = en_US

\section{Conclusion}
\label{sec:conclusion}

In this paper, we developed a new MITM system for modern cellular network protocols based on overshadowing and passive sniffing, called AdaptOver. Previously, higher-layer protocol attacks between a UE and the core network that allowed long-term DoS and identity leakage required the use of a fake base station. Using AdaptOver, we removed this assumption by implementing message overshadowing and injection. At the same time, using uplink overshadowing, AdaptOver is much more efficient in terms of required power and achieved range. We showed how an attacker can use AdaptOver to launch either a large-scale DoS/privacy-leaking campaign or a targeted attack based on a phone number. In collaboration with an operator, we showcased the capabilities of this system in the real world by attacking a user 3.8km away from the attacker. Finally, we presented potential countermeasures for baseband and radio network equipment vendors.

%% file: sections/91_acks.tex
% !TeX spellcheck = en_US

%% ERC Grant
This project has received funding from the European Research Council (ERC) under the European Union’s Horizon 2020 research and innovation program under grant agreement No 726227.
%% NCCR
Research supported by the Swiss National Science Foundation under NCCR Automation, grant agreement 51NF40\_180545.